# HEAVY BARYON SPECTROSCOPY

UKQCD Collaboration, Talk presented by N. Stella[a]

[a]Department of Physics, The University, SO17 1BJ Southampton, UK

We present the preliminary results of an exploratory study of heavy baryon spectroscopy, using the $O(a)$-improved fermionic action. Estimates of masses and splittings at the charm and beauty physical limit are reported.

The study of heavy flavours on the lattice is proving to be a fruitful field of investigation, although, so far, it has been restricted to the mesonic sector [1]. We have recently begun to explore the possibility of studying Heavy Baryon spectroscopy, which is not only interesting *per se*, but is also a necessary step in the computation of baryonic weak matrix elements. Only baryons containing a single heavy quark have been considered. They are listed in tab. 2, together with their quantum numbers and those of their light degrees of freedom. We have computed the correlation function of the operator $\mathcal{O}_5 = \epsilon_{abc}(l^a \mathcal{C} \gamma_5 l'^b) h^c$, where $l, l', h$ are the fields of the light and the heavy quarks respectively, and $\mathcal{C}$ is the charge conjugation matrix. At large $t$ and for $t < T/2$, the correlation function

$$G_5(\vec{0}, t) = \sum_{\vec{x}} \langle \mathcal{O}_5(\vec{x},t) \overline{\mathcal{O}}_5(\vec{0},0) \rangle \to Z_5 \frac{\gamma_0 + 1}{2} e^{-M_5 t}$$

gives the mass of the $\Lambda$ and $\Xi$ baryons. The operator $\mathcal{O}_\mu = \epsilon_{abc}(l^a \mathcal{C} \gamma_\mu l'^b) h^c$ has also been considered. Its 2-pt function

$$G_{\mu\nu}(\vec{0}, t) = \sum_{\vec{x}} \langle \mathcal{O}_\mu(\vec{x},t) \overline{\mathcal{O}}_\nu(\vec{0},0) \rangle \to$$

$$Z_{3/2}^2 e^{-tM_{\frac{3}{2}}} \frac{\gamma_0 + 1}{2} P^{3/2}_{\mu\nu} + Z_{1/2}^2 e^{-tM_{\frac{1}{2}}} \frac{\gamma_0 + 1}{2} P^{1/2}_{\mu\nu}$$

is given by the superposition of the spin $s = 1/2$ and $s = 3/2$ contributions. At rest, the signals of the spin doublets $(\Sigma, \Sigma^*)$, $(\Xi', \Xi^*)$ and $(\Omega, \Omega^*)$ can be separated using the two projectors [2]

$$P^{3/2}_{ij} = g_{ij} - \frac{1}{3}\gamma_i\gamma_j; \quad P^{1/2}_{ij} = \frac{1}{3}\gamma_i\gamma_j, \; i,j = 1,2,3.$$

The unphysical degrees of freedom, present since $\partial_\mu \mathcal{O}_\mu \neq 0$, contribute only to $G_{00}(\vec{0}, t)$. We have generated gauge configurations and quark propagators with the parameters presented in tab. 1, adopting the Jacobi [3] smearing method, at the source (SL) or at both the source and the sink (SS). From the fits we have evidence that the SS 2-pt functions are less correlated and give more stable results. A typical fit to the $\Lambda$, see fig. 1, would give an error of about 1% on the mass and a correlated $\chi^2/\mathrm{dof} = 0.8$. Similarly, for the $\Sigma$ we find $\chi^2/\mathrm{dof} = 1.0$, though the mass is measured with a slightly larger error. Hence, all the numbers quoted in the following have been obtained by fitting the SS correlators, in $t \in [15, 23]$. In this exploratory study only a restricted set of quark masses has been considered. Therefore, extrapolations have been modelled by simple linear behaviours both in the light quark masses, to the chiral and strange limit, and in the heavy quark mass to the charm and beauty limit. In fig. 2 we show that the extrapolations are very smooth.

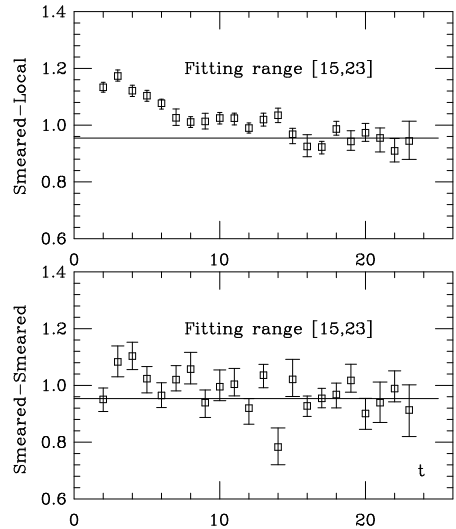

Figure 1. Effective mass for the $\Lambda$ baryon, at $\kappa_h = 0.129$ and $\kappa_{l1} = \kappa_{l2} = 0.14144$. Comparison between SL and SS cases.

The measured masses, presented in tab. 2, depend linearly on the lattice spacing and they include an implied uncertainty represented by the factor $a^{-1}/(2.7\,\mathrm{GeV})$ The charmed baryon masses compare well with the experimental measurements. The beautiful ones, on the other hand, are affected by large errors. This is due to the fact that the heavy extrapolation relies on the only two $\kappa_h$ values we have computed so far, which lie far away from the b-quark limit. The extraction of mass splittings is delicate since they decrease with the heavy quark mass and, in this study, are expected to have the same size (few tens of MeV) as our statistical errors. The splittings seem to be very sensitive to the extrapolation procedure, therefore we can only achieve semiquantitative results for the beauty sector.

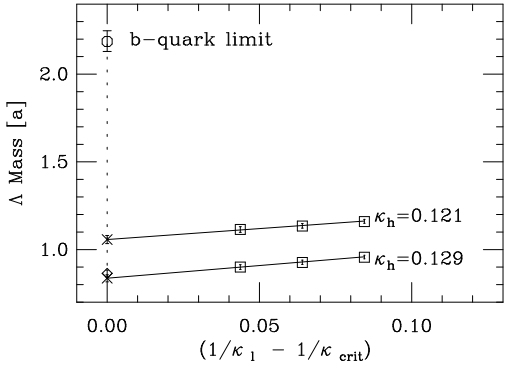

Figure 2. Extrapolations of the $\Lambda$ baryon, to the chiral limit (solid line) and to the heavy quark mass (dashed line).

The $M_\Lambda - M_P$ splitting has been computed as a function of the pseudoscalar mass $M_P$, and ex-

| Dataset | Action | Light $\kappa$ | Heavy $\kappa$ |
|---|---|---|---|
| 60 cfgs | Improved | 0.14144/14144 | 0.121 |
| $24^3 \times 48$ | refs. [5,6] | 0.14144/14226 | 0.129 |
| $\beta = 6.2$ | | 0.14226/14226 | |

Table 1
Parameters adopted in the present simulation.

trapolated linearly:

$$(M_\Lambda - M_P) = A + \frac{B}{M_P}.$$

To estimate the slope with the smallest possible uncertainty, we have fitted the double ratio of the correlation functions of the $\Lambda$, $\langle \Lambda(\kappa_h) \rangle$ and of the pseudoscalar $\langle P(\kappa_h) \rangle$ at different values of $\kappa_h$:

$$\left.\frac{\langle \Lambda(\kappa_h=0.121)\rangle \langle P(\kappa_h=0.129)\rangle}{\langle \Lambda(\kappa_h=0.129)\rangle \langle P(\kappa_h=0.121)\rangle}\right|_{\kappa_l} \sim e^{(-\Delta(M_\Lambda - M_P)t)}.$$

The increment of the splitting $\Delta(M_\Lambda - M_P)$ is extracted from the previous expression, in which some of the fluctuations are expected to cancel, This way, at fixed value of the light quark mass $\kappa_l$, we compute $B = \Delta(M_\Lambda - M_P)/\Delta(M_P^{-1})$ and we then extract $A$ by fitting the single ratio

$$\left.\frac{\langle \Lambda(\kappa_h)\rangle}{\langle P(\kappa_h)\rangle}\right|_{\kappa_l} \sim \exp\left(-(M_\Lambda - M_P)t\right).$$

Having obtained $A$ and $B$ at fixed values of the light quark masses, we extrapolate the result to the chiral limit. In fig. 3, we plot the splitting at fixed $\kappa_h$, at the chiral limit, computed by considering the difference $M_\Lambda - M_P$, and the extrapolation to the physical limit, obtained by using our estimates of $B$ and $A$. We note that the extrapolation of the splitting reproduces remarkably well the UKQCD static point [7]. In ref. [8], the values of the splittings obtained with Wilson fermions are reported and compared with ours.

We would like to mention that a similar analysis for the $M_\Lambda - M^{\mathrm{av}}$, where by $M^{\mathrm{av}}$ we mean the spin-averaged meson mass, has been performed. In this case the discrepancy between lattice estimates and experimental values is slightly larger. Our estimates for the splitting at the physical $c$- and $b$-quark masses are reported in tab. 3. We prefer to compute the ratio of the splittings to the sum of the masses, to deal with dimensionless quantities. The $M_\Sigma - M_\Lambda$ splitting, see also tab. 3 has been analyzed adopting the same procedure. Although, in this case, the size of the error bars greatly reduces the quantitative interpretation.

Finally, we report results on the mass splittings of the spin doublets, at the charm mass. $M_{\Sigma^*} - M_\Sigma$ appears underestimated with respect to experimental determination. With the present error bars, we cannot tell if this is a statistical effect or rather a systematic one. If the latter is the case, this situation resembles the one which is well known in the case of the pseudoscalar-vector splitting, see refs. [9,10].

| Baryon | $J^P$ | $(I)(S)s_l^{\pi_l}$ | Quark Content | h = charm Exp. | h = charm Latt. | h = beauty Exp. | h = beauty Latt. |
|---|---|---|---|---|---|---|---|
| $\Lambda_h$ | $\frac{1}{2}^+$ | $(0)(0)\ 0^+$ | $(ud)h$ | 2285 | $2334\ ^{+48}_{-57}$ | 5641 | $5900\ ^{+170}_{-150}$ |
| $\Sigma_h$ | $\frac{1}{2}^+$ | $(1)(0)\ 1^+$ | $(uu)h$ | 2453 | $2498\ ^{+82}_{-83}$ | – | $5710\ ^{+250}_{-170}$ |
| $\Sigma_h^*$ | $\frac{3}{2}^+$ | $(1)(0)\ 1^+$ | $(uu)h$ | 2530 | $2443\ ^{+90}_{-61}$ | – | $5630\ ^{+220}_{-180}$ |
| $\Xi_h$ | $\frac{1}{2}^+$ | $(\frac{1}{2})(-1)\ 0^+$ | $(us)h$ | 2469 | $2463\ ^{+40}_{-51}$ | – | $6000\ ^{+130}_{-150}$ |
| $\Xi_h'$ | $\frac{1}{2}^+$ | $(\frac{1}{2})(-1)\ 1^+$ | $(us)h$ | – | $2609\ ^{+71}_{-73}$ | – | $5870\ ^{+200}_{-180}$ |
| $\Xi_h^*$ | $\frac{3}{2}^+$ | $(\frac{1}{2})(-1)\ 1^+$ | $(us)h$ | – | $2579\ ^{+80}_{-80}$ | – | $5829\ ^{+170}_{-160}$ |
| $\Omega_h$ | $\frac{1}{2}^+$ | $(0)(-2)\ 1^+$ | $(ss)h$ | 2740 | $2688\ ^{+47}_{-38}$ | – | $5880\ ^{+130}_{-70}$ |
| $\Omega_h^*$ | $\frac{3}{2}^+$ | $(0)(-2)\ 1^+$ | $(ss)h$ | – | $2654\ ^{+52}_{-33}$ | – | $5870\ ^{+130}_{-70}$ |

Table 2
Heavy Baryons considered in this project. We have listed the relevant quantum numbers of each baryon and those of its light degrees of freedom. Our determination and the available experimental data are reported.

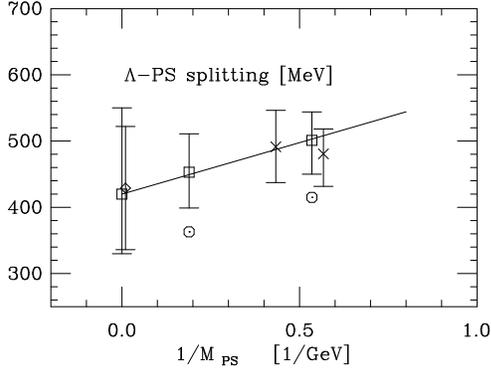

Figure 3. $M_\Lambda - M_P$ mass splitting, at the actual $\kappa_h$ considered ($\times$), and at the physical masses ($\square$). The UKQCD static point ($\diamond$), and the experimental determinations ($\odot$) are also shown.

A lattice study of heavy baryons is certainly feasible. We get good signals for the correlation functions, even if the size of the errors and the systematic uncertainties, call for a better control of the extrapolations, in particular for small quantities, like mass splittings. However, an accurate measure of mass splittings is not a necessary precondition for studying weak matrix elements.

|  | $(M_{\Lambda_c}-M_D)/(M_{\Lambda_c}+M_D)$ | $(M_{\Lambda_b}-M_B)/(M_{\Lambda_b}+M_B)$ |
|---|---|---|
| double | $0.120\ ^{+10}_{-11}$ | $0.037\ ^{+15}_{-10}$ |
| Exp. | 0.100 | 0.033 |

|  | $(M_{\Sigma_c}-M_{\Lambda_c})/(M_{\Sigma_c}+M_{\Lambda_c})$ | $(M_{\Sigma_b}-M_{\Lambda_b})/(M_{\Sigma_b}+M_{\Lambda_b})$ |
|---|---|---|
| double | $0.038\ ^{+52}_{-46}$ | $0.017\ ^{+13}_{-13}$ |
| Exp. | 0.035 | - |

|  | $M_{\Sigma_c^*}-M_{\Sigma_c}$ | $M_{\Xi_c^*}-M_{\Xi_c'}$ | $M_{\Omega_c^*}-M_{\Omega_c}$ |
|---|---|---|---|
| ratio [MeV] | $-50\ ^{+62}_{-60}$ | $-27\ ^{+32}_{-35}$ | $-31\ ^{+40}_{-26}$ |
| Exp. | 61 | - | - |

Table 3
Summary of results for several mass splittings.